# Prediction of Stable Ground-State Lithium Polyhydrides under High Pressures

Yangmei Chen,[1,2] Hua Y. Geng,[1,*] Xiaozhen Yan,[1,2,3] Yi Sun,[1]

Qiang Wu,[1,*] and Xiangrong Chen[2,*]

[1]National Key Laboratory of Shock Wave and Detonation Physics, Institute of Fluid Physics, CAEP; P.O. Box 919-102, Mianyang, Sichuan, People's Republic of China, 621900

[2]Institute of Atomic and Molecular Physics, College of Physical Science and Technology, Sichuan University, Chengdu, Sichuan, People's Republic of China, 610065

[3]School of Science, Jiangxi University of Science and Technology, Ganzhou, Jiangxi, People's Republic of China, 341000

*Correspondence should be addressed to Hua Y. Geng, Qiang Wu, and Xiangrong Chen (s102genghy@caep.cn, wuqianglsd@163.com, xrchen@scu.edu.cn)

**Abstract**

Hydrogen-rich compounds are important for understanding the dissociation of dense molecular hydrogen, as well as searching for room temperature Bardeen-Cooper-Schrieffer (BCS) superconductors. A recent high pressure experiment reported the successful synthesis of novel insulating lithium polyhydrides when above 130 GPa. However, the results are in sharp contrast to previous theoretical prediction by PBE functional that around this pressure range all lithium polyhydrides (LiH$_n$ ($n$ = 2-8)) should be metallic. In order to address this discrepancy, we perform unbiased structure search with first principles calculation by including the van der Waals interaction that was ignored in previous prediction to predict the high pressure stable structures of LiH$_n$ ($n$ = 2-11, 13) up to 200 GPa. We reproduce the previously predicted structures, and further find novel compositions that adopt more stable structures. The van der Waals functional (vdW-DF) significantly alters the relative stability of lithium polyhydrides, and predicts that the stable stoichiometries for the ground-state should be LiH$_2$ and LiH$_9$ at 130-170 GPa, and LiH$_2$, LiH$_8$ and LiH$_{10}$ at 180-200 GPa. Accurate electronic structure calculation with GW approximation indicates that LiH, LiH$_2$, LiH$_7$, and LiH$_9$ are insulative up to at least 208 GPa, and all other lithium polyhydrides are metallic. The calculated vibron frequencies of these insulating phases are also in accordance with the experimental infrared (IR) data. This reconciliation with the experimental observation suggests that LiH$_2$, LiH$_7$, and LiH$_9$ are the possible candidates for lithium polyhydrides synthesized in that experiment. Our results reinstate the credibility of density functional theory in description H-rich compounds, and demonstrate the importance of considering van der Waals interaction in this class of materials.

**Introduction**

High pressure stability of hydrogen-rich compounds has attracted a lot of attention because of the richness of these materials in physics and chemistry that has been revealed by recent theoretical and experimental works[1-18]. Some of them (*e.g.*, $H_3S$ was predicted theoretically and then confirmed by experiment) would be possible candidates of "high temperature" Bardeen-Cooper-Schrieffer (BCS) superconductors[18-21], and some others (*e.g.*, $LiH_n$) have important applications in the field of hydrogen storage and energy industry. Direct laboratory synthesis of hydrogen rich compound $LiH_n$ ($n > 1$) is difficult, and may be hindered by kinetic barriers when LiH and $H_2$ are mixed together. It is well known that LiH keeps stable up to 160 GPa at 300 K, and has no chemical reaction with the molecular hydrogen[22]. Using laser to heat LiH+$H_2$ mixture in a diamond anvil cell (DAC) is helpful to overcome this barrier, and this method provided the first preliminary experimental evidence of the formation of lithium polyhydrides that containing $H_2$ units (referred to as $LiH_x$-II). It indicated that $LiH_x$-II was synthesized at 5 GPa and 1800 K, which is transparent up to 62 GPa[23]. Nevertheless, possible chemical contamination is difficult to avoid in the heating process. Also, a recent similar experiment up to 50 GPa and 2000 K was not able to detect any polyhydrides of Li[24]. In that experiment, however, novel $NaH_3$ and $NaH_7$ have been successfully synthesized. This indicates higher pressure is required for $LiH_n$ synthesis. Indeed, it was reported recently that at 300 K the lithium polyhydrides containing $H_2$ units have also been synthesized when above 130 GPa. The new phases were detected using a noninvasive probing technique of synchrotron infrared (IR) absorption that has a high sensitivity to chemical changes[25]. The experiment unequivocally showed that these polyhydrides remain insulating up to 215 GPa, and the shifts of vibron IR frequencies exhibit a positive pressure dependence. This experiment was inspired by previous prediction made using the density functional theory on the stability of $LiH_n$ ($n$ = 2-8) with the Perdew-Burke-Enzerhof (PBE) functional[26]. The results of the theoretical calculation and experiment are consistent in the formation pressure of lithium polyhydrides. And the measured vibron frequencies also roughly coincide with the most intense frequencies calculated for $LiH_2$ and $LiH_6$. It was tentatively assumed that the synthesized polyhydrides are $LiH_6$ and $LiH_2$, which contain $H_2$ units in the structure, unfortunately, the observed insulative character contradicts with the predicted metallic behavior of these phases sharply. This qualitative discrepancy stimulates us to pursue further theoretical investigations on LiH$_n$ with more accurate exchange-correlation functionals.

It is well known that the results of density functional theory (DFT) depend on the exchange-correlation functional[27]. Most previous structure searching and electronic structure calculations for hydrogen and alkali and alkaline earth hydrides were performed using the semi-local generalized gradient approximation (GGA), usually in the parameterization of PBE. It gives a good description of the structural stability of lithium and sodium, as well as the melting curve of hydrogen[28-31]. However, it was found very recently that PBE is poor when describing the dissociation of H$_2$ molecules, and the correction from the van der Waals interaction is essential[32], which was demonstrated by comparison to the most accurate quantum Monte Carlo (QMC) calculation, and the vdW-DF functional was shown to provide the most consistent bond length of the molecular phase of hydrogen with respect to QMC[32]. Since many hydrogen-rich compounds also contain H$_2$ units, and some even involve H$_2$ dissociation, it is expected that van der Waals interaction might play a significant role in the structural stability of these compounds. But this contribution was completely ignored in most previous DFT calculations, which could be the origination of the discrepancy between the theory and experiment on lithium polyhydrides.

In this work, we combine the unbiased structural search using the particle-swarm optimization (PSO) algorithm[33, 34] and DFT calculation with the van der Waals functional (vdW-DF)[35] to systematically investigate the phase stability of lithium polyhydrides LiH$_n$ ($n$ = 2-11, 13) under high pressures. Our results reveal that the van der Waals interaction indeed alters the relative stability of lithium polyhydrides greatly. Furthermore, the electronic structure and vibrational property are also modified with respect to the PBE functional. The compounds with stoichiometry of LiH, LiH$_2$, LiH$_7$ and LiH$_9$ are predicted to have a band gap up to at least 208 GPa, which agrees well with the observed insulative characters in the experiment. The pressure dependence of their vibron frequencies also has a trend similar to the experimental data[25]. These results constitute a much better theoretical interpretation of the recent experimental data.

**Computational Details**

We searched for high pressure structures of lithium polyhydrides using the particle swarm optimization methodology as implemented in the CALYPSO code[33, 34]. This approach is unbiased by any prior known structure information and has demonstrated good efficiency in predicting high pressure structures of hydrogen-rich compounds[7-9, 15, 16, 36] and other novel materials[37-41]. The size of the system in our structure searches contains 1-4 formula units per simulation cell. Each search generation contains 30-50 structures and the structure searching simulation is usually stopped after generates 900-1500 structures. The underlying *ab initio* total energy calculations and structural relaxations are carried out using the plane wave basis and projector-augmented-wave (PAW) method[42] with the non-local dispersion corrected density functional of Dion *et al.* (vdW-DF)[35] as implemented in the Vienna ab-initio simulation package (VASP)[43]. The use of a cutoff energy of 950 eV and dense enough k-point sampling grids give excellent convergence of the calculated enthalpy (<1 meV/atom). The atomic charges are obtained from Bader topological analysis[44-46] using denser grids at a high level of accuracy. Lattice dynamics and phonon contributions are calculated to verify the dynamical stability of predicted phases by the small displacement method as implemented in the PHONOPY package[47].

**Results and Discussion**

1. **Structure and convex hull**

We perform structure searches for $LiH_n$ ($n$ = 2-11, 13) at pressures of 150 and 200 GPa using vdW-DF functional, respectively. Thermodynamic stabilities of the $LiH_n$ compounds under various pressures are then investigated by calculating the formation enthalpy ($\Delta H$) with respect to decomposition into LiH and $H_2$, as shown in Figures 1 and 2. It is necessary to point out that LiH is always the stable compound of $LiH_n$ when n<2[26], therefore it is not necessary to study the stability of $LiH_n$ with respect to the constituent elements Li and $H_2$. Our vdW-DF calculations predict that the stable stoichiometries of polyhydrides are $LiH_2$ and $LiH_9$ at 130-170 GPa, which then shift to $LiH_2$, $LiH_8$, and $LiH_{10}$ at 180-200 GPa. $LiH_2$ has the most negative formation enthalpy with respect to LiH and $H_2$ within the whole pressure range we studied. These accurate results are in sharp contrast to the previous PBE predictions of $LiH_n$ ($n$ = 2-8)[26], which suggested that $LiH_2$, $LiH_6$, and $LiH_8$ are the three most stable lithium polyhydrides at pressures between 150-200 GPa, and $LiH_6$ was predicted to be the most stable one. The later phase, however, is always metastable at the level of vdW-DF. Additionally, we also recalculate the formation enthalpies of $LiH_{16}$ with vdW-DF functional using the structure given in Ref. [48] that was predicted by PBE functional, and find that it shifts up to above the convex hull when the van der Waals interaction is included, indicating that LiH$_{16}$ is actually unstable or metastable in energetics. A comparison of the phase stability of LiH$_n$ calculated by LDA, PBE, and vdW-DF is presented in Figure 1. It can be clearly seen that the van der Waals interaction has a dramatic impact on the stability of LiH$_n$: It not only adds two new phases LiH$_9$ and LiH$_{10}$ to the ground states (their structures are presented in Figure 3), but also removes LiH$_6$ away from the convex hull. This effect of van der Waals interaction is a combination result of the better description of the electronic structure by advanced vdW-DF functional and the subsequent modification on the atomic structures. Table S1 in the Supporting Information (SI) provides the shortest bond length (H-H, Li-H and Li-Li) in lithium polyhydrides and their relative deviations of PBE with respect to the results calculated by vdW-DF functional. It is found that the vdW-DF corrections are mainly on the H-H distance that leads to shorter H$_2$ bond length, and this is fully in line with our initial expectation. The consequence is that the van der Waals correction leads to slightly larger lattice parameters (see Table S3 in the Supporting Information), and the formation enthalpies are significantly reduced, thus alters the relative phase stability of lithium polyhydrides greatly. Inclusion of the zero-point energy at 0 K or the finite temperature free energy at 300 K has a negligible influence on the convex hull (see Figure S2 in the Supporting Information).

It should be mentioned that although the vdW-DF functional[35] significantly modifies the relative stabilities of LiH$_2$, LiH$_6$, and LiH$_8$, the re-optimized structures of these phases using vdW-DF still have the same space groups as those obtained by PBE[26]. The structure details of them are presented in Table S4 of the Supporting Information. In addition, it should be emphasized that it is the first time that LiH$_9$ and LiH$_{10}$ phases are predicted to be stable ground states in lithium polyhydrides. The newly discovered LiH$_9$ has three distinct structures, each with two formula units per cell (see Figure 3b): the first one (space group *Cmc*2$_1$) stabilizes at 150-196 GPa, the second one (space group *Cc*) is stable up to 223 GPa, and then transforms to the third one (with space group *P*-1) when beyond 223 GPa. LiH$_{10}$ stabilizes in a structure of *C*2/*c* with two formula units per cell. In order to examine the dynamical stability of the predicted LiH$_n$ compounds, lattice dynamics are calculated with quasi-harmonic approximation. The results show that all predicted structures have no imaginary phonon modes, except for LiH$_{11}$ that is not on the convex hull. This indicates that these predicted compounds should be stable or at least be meta-stable, and could be detected experimentally. In particular, LiH$_7$ and LiH$_9$ are meta-stable at 180-200 GPa, where they have a slightly higher formation enthalpy with respect to LiH$_2$, LiH$_8$, and LiH$_{10}$ at this pressure range.

The structural elements in these stable or meta-stable phases can be divided into three categories (see Figure S3 in the Supporting Information): (*i*) Li$^+$ ions and H$_2$ dimers; (*ii*) Li$^+$ ions, H$_2$ dimers, and hydridic H$^-$ anions; and (*iii*) Li$^+$ ions, H$_2$ dimers, and asymmetric H$_3^-$ clusters. Only the first two cases were observed in the previous PBE calculations of lithium polyhydrides[26], which appear in LiH$_2$, LiH$_6$, and LiH$_8$, respectively. The newly discovered phases LiH$_7$ and LiH$_9$ also contain asymmetric H$_3^-$ units. This is analogous to the newly synthesized NaH$_3$ and NaH$_7$, where the former contains H$_2$ units and the latter also has H$_3^-$ motif[24]. Both phases are insulative, similar to the experimentally observed LiH$_n$[25]. The observation of asymmertric H$_3^-$ anions is astonishing since it was believed that such motif cannot occur in LiH$_n$[49], though they are ubiquitous in heavy alkali polyhydrides[49-51]. As shown in Figure 1, the vdW-DF functional brings the LiH$_7$ and LiH$_9$ phases closer to the convex hull, in which the LiH$_9$ phase locates exactly on the convex hull in comparison with those predicted by PBE, respectively. For lithium-rich hydrides and heavy alkali polyhydrides, van der Waals interaction might also affect their relative stabilities. For this reason, we also employ the vdW-DF functional to recalculate the formation enthalpies for the previously predicted structures of Li$_n$H (*n*=3-9) at 90 GPa[52] and NaH$_x$ (*x* =3, 6-12) at 50, 100, and 300 GPa[24, 53], respectively (see Figures S4 and S5 in the Supporting Information). It is found that the vdW-DF corrections in NaH$_x$ are smaller than those in LiH$_n$, and keep their relative stabilities unchanged in comparison with those predicted by PBE. This is likely due to the stronger coulomb interaction between Na and H arising from the lower ionization potential and larger ionic radius of Na as compared with Li[53]. But in lithium polyhydrides, the van der Waals interaction favors the formation of H$_3^-$ units when hydrogen content is high. Nevertheless, higher compression eventually turns LiH$_7$ and LiH$_9$ into those structures that containing only H$_2$ units, showing the narrow stability range of H$_3^-$ units in lithium polyhydrides. In other words, for LiH$_n$ at this pressure range, the amount of charges donated by lithium ions are not sufficient to dissociate all H$_2$ dimers completely, and the van der Waals interaction enhances the localization of these charges onto some H atoms and H$_2$ dimers to form H$^-$ and H$_2^{\delta-}$ anions. The resultant increased H$_2$...H$^-$ interaction then leads to the formation of H$_3^-$ clusters. At higher pressures, however, these transferred charges delocalize further and spread out over the whole cell (becoming metallic), thus alleviate the $H_2$ dissociation, and $H_3^-$ units become unfavorable again. This mechanism is different from what observed in heavy alkali polyhydrides. Though in both cases the appearance of $H_3^-$ units is driven by the tendency to form multi-centered bonding[54], here it is preferred because of the enhanced van der Waals interaction between $H^-$ anions and $H_2$ dimers, whereas it is supposed to be because of the softness of the cations in the heavy alkali polyhydrides[49-51].

**2. Electronic properties**

The recent high-pressure experiment revealed an insulating state in the synthesized lithium polyhydrides up to at least 215 GPa[25]. This is quite unexpected and contradicts the previous theoretical predictions made by PBE, which suggested that all phases of $LiH_n$ ($n$ = 2-8) are metallic from 100 to 300 GPa[26]. One possible reason of this discrepancy is that some stable or meta-stable structures have been omitted in previous structure searches, as our newly discovered $LiH_7$ and $LiH_9$ suggested. Using the vdW-DF functional, the electronic properties of the newly predicted $LiH_n$ are also investigated by calculating their electronic density of states and band structures. The results reveal that LiH, $LiH_7$, and $LiH_9$ are all insulators with a band gap of 2.21, 1.06, and 2.14 eV at 150 GPa, respectively, whereas other lithium polyhydrides exhibit metallic feature. The pressure dependence of the band gaps are given in Figure 5, wherein all of the band gaps decrease with increasing pressure. LiH was reported to keep the B1 phase up to 300 GPa[55, 56], and our vdW-DF calculations show that it remains an insulator with a band gap of 1.01 eV at this pressure. The meta-stable $LiH_7$ is predicted to undergo a pressure-induced structural transition at about 158 GPa. Both phases of $LiH_7$ have the same space group of *P*-1 (Figure 4a). At 150 GPa, their band gaps are 1.06 and 1.05 eV, which reduce to 0.30 and 0.64 eV at 220 GPa, respectively (Figure 5c). As mentioned above, $LiH_9$ has two sub-sequent phase transitions with three distinct structures within the pressure range of 150-230 GPa. The former two are predicted to be insulative up to at least 220 GPa (Figure 5d), and the last one is metallic within its stable pressure range.

It should be noted that the vdW-DF and PBE functionals produce different energy band structures and density of states (see Figure S6 in the Supporting Information, taking $LiH_2$ as an example, other cases of $LiH_n$ are similar to that of $LiH_2$) determined by the distribution of the electron wave function boiling down to their bond length, which leads to different values for the band gap. The vdW-DF functional predicts that the band gap of LiH$_2$ closes up at around 150 GPa, whereas the PBE predicts a metallization pressure of 50 GPa[57]. At the vicinity of the Fermi level, the contribution to the density of states arises mainly from H$^-$ or H$_2$ bonding σ states and H$_2$ anti-bonding σ* states in the occupied valence bands and unoccupied conduction bands, respectively, as illustrated in Figure S7 in the Supporting Information. Specifically, adding additional hydrogen into LiH leads to charge transfer from lithium cations to H$_2$ dimers and removes electrons from Li 3$d$ states[55]. But in LiH$_2$ there still are Li 2$p$ and Li 2$s$ states that are partially occupied. Hybridization of them with H$^-$ 1$s$ and H$_2$ 1s states opens the band gap. Further addition of hydrogen removes electrons from Li 2$p$ and Li 2$s$ states and makes them unoccupied, they hybridize with H$_2$ 1$s$ only in the conduction bands. This, however, cannot open up the band gap, and just leads to a deep dip in the density of states near the Fermi level (see LiH$_6$, LiH$_8$, LiH$_9$ (P-1), and LiH$_{10}$). Formation of H$_3^-$ clusters in LiH$_7$ and LiH$_9$ (that is further stabilized by the van der Waals interaction) leads to H$_3^-$ 1$s$ and H$_2$ 1$s$ hybridization, which eventually opens the band gaps, and the resultant empty states further hybridize with the unoccupied Li 2$p$ and H$_2$ 2$p$ states. It is evident that in H-rich lithium polyhydrides, emergence of H$_3^-$ clusters is the key to open up the band gap and to become insulative. In this respect, the bulk of the synthesized LiH$_n$ must be one of LiH$_7$ and LiH$_9$. In fact, both PBE and vdW-DF functionals underestimate the band gap. In order to obtain an accurate estimation of the band gaps, we compute the band gap as a function of pressure by using the GW approximation[58-61], for which the vdW-DF wave functions and eigenvalues are used as the initial guess (vdW-DF+GW). As shown in Figure 5, it is found that pure vdW-DF calculations underestimate the band gaps of LiH, LiH$_2$, LiH$_7$, and LiH$_9$ by about 0.80-1.65 eV by comparison with the GW method. Moreover, the metallization pressure of LiH$_2$ obtained by vdW-DF+GW is increased to 208 GPa, which is much higher than the PBE+GW result of 170 GPa[57], revealing the importance of van der Waals interaction to produce a good initial wave function for GW calculation. Up to at least 200 GPa, the vdW-DF+GW calculations show that all of LiH, LiH$_2$, LiH$_7$, and LiH$_9$ are insulative, in which both LiH and LiH$_2$ are the ground states, and LiH$_9$ is stable between 130-170 GPa and becomes metastable beyond that pressure. This is qualitatively in accordance with the recent experimental observation in the synthesized lithium polyhydrides[25], and suggests that insulative LiH$_n$ ($n$>2) must contain H$_3^-$ clusters.

### 3. Vibron frequencies

The vibron frequencies of LiH$_2$ and LiH$_6$ in previous PBE calculations for LiH$_n$ ($n$ = 2-8)[26] were lately employed to lithium polyhydrides, and matched the infrared spectra observed in the experiment roughly[25]. But the poor quality of PBE when describing the structures and energetics of lithium polyhydrides greatly undermines its credibility, not to mention that the predicted metallic behavior of these two phases is completely in disagreement with the insulating feature up to 215 GPa observed in the experiment[25]. Based on our predicted phase stability of insulating LiH (*Fm-3m*), LiH$_2$ (*P$_4$/mbm*), LiH$_7$ (*P*-1(2)), and LiH$_9$ (*Cmc*2$_1$ and *Cc*) in LiH$_n$ ($n$ = 2-11, 13), we also employ the vdW-DF functional to compute their vibron frequencies, and compare them to the experimental infrared (IR) spectra data. The vibron frequencies of metallic LiH$_6$ (*R-3m*), LiH$_8$ (*I*422) and LiH$_{10}$ (*C*2/*c*) are also included for comparison. As illustrated in Figure 6, the calculated vibron frequencies of LiH are in good agreement with the experimental LO and TO modes[62], showing the adequacy of our method. It also signals that the higher vibron modes must come from other polyhydrides. In the case of LiH$_2$, one of its IR active vibron frequencies is roughly consistent with the experimental data of the mode $\nu_3$ at high pressures. The vibron frequencies of LiH$_7$ and LiH$_9$ (*Cmc*2$_1$) are slightly larger than those of LiH$_2$ in the whole pressure range considered here. For all of the predicted phases, the experimentally measured $\nu_3$ mode is best described by LiH$_9$ and in the *Cc* structure. However, none of the predicted insulator phases have a vibron frequency that can match the observed $\nu_1$ and $\nu_2$ modes. For metallic LiH$_6$, LiH$_8$, and LiH$_{10}$ phases, their vibron frequencies predicted by vdW-DF functional are larger than the experimental data of $\nu_1$ and $\nu_2$ modes, and much lower than the experimental value of $\nu_3$ mode.

It should be noted that previous calculations indicated that the vibron frequencies of solid H$_2$ (in the phases of *P*6$_3$/*m* and *C*2/*c*) decrease with increasing pressure[63], and the high-frequency regime (>3500 cm$^{-1}$) depends strongly on the choice of exchange-correlation functional[64]. In order to examine the effect of different exchange-correlation functional on the vibron frequencies of LiH$_n$, we also calculate the vibron frequencies of the above mentioned polyhydrides using the PBE functional. It is found that PBE shifts down the vibron frequency by approximately 500 cm$^{-1}$ for LiH$_2$, 380 cm$^{-1}$ for LiH$_6$, 310 cm$^{-1}$ for LiH$_7$, 180 cm$^{-1}$ for LiH$_8$, 350 cm$^{-1}$ for LiH$_9$, and 240 cm$^{-1}$ for LiH$_{10}$. All of these structures contain H$_2$ or H$_3^-$ units. On the other hand, for LiH that has no H$_2$ or H$_3^-$ units, the PBE results agree well with those of vdW-DF (see Figure S8 in the Supporting Information). This analysis indicates that the van der Waals interaction is important and significantly influences the vibrational properties in the range of high-frequency (>2500 cm$^{-1}$) of lithium polyhydrides that containing $H_2$ or $H_3^-$ units, which is consistent with the observation that vdW-DF corrections mainly impact on the H-H bond length. The result clearly shows that the van der Waals correction removes the previous agreement of the vibron frequency of $LiH_6$ with the experimental data suggested by PBE calculations. It also should be noted that in solid $H_2$, anharmonicity and finite temperature effects shift the vibron frequencies at about several hundred cm$^{-1}$. This correction magnitude is at the same level of the deviation between the vibron frequencies of $LiH_6$ and $LiH_8$ and the experimental $\nu_1$ and $\nu_2$ data. Thus these two modes might contain contribution from the meta-stable $LiH_6$ or $LiH_8$. The total amount of them, however, must be very tiny so that has negligible influence on the overall transparency of the reacted sample. Another possibility of the discrepancy might be that the experimentally observed $\nu_1$ and $\nu_2$ IR frequencies come from the influence of carbon (of DAC) on the vibron properties of $LiH_n$, which is still a topic requires further exploration. According to above predicted insulative characters and vibron frequencies of lithium polyhydrides with vdW-DF calculations, we conclude that the possible candidates for the majority of the lithium polyhydrides that were synthesized in the recent experiment should be $LiH_2$, $LiH_9$, or the meta-stable $LiH_7$, with $LiH_9$ has the highest possibility.

**Conclusions**

The phase stability of $LiH_n$ ($n$ = 2-11, 13) at high pressures are investigated using the particle-swarm optimization algorithm in combination with first-principles calculations using vdW-DF functional to take the van der Waals interaction into account. We found that the van der Waals correction significantly alters the relative stabilities of lithium polyhydrides, predicting that $LiH_2$ and $LiH_9$ are stable at 130-170 GPa, and $LiH_2$, $LiH_8$, and $LiH_{10}$ become stable at 180-200 GPa. Most importantly, we discovered that besides LiH, polyhydrides $LiH_2$, $LiH_7$, and $LiH_9$ also keep the insulating feature up to at least 208 GPa, which is in sharp contrast to the metallic behavior of $LiH_2$, $LiH_6$, and $LiH_8$ predicted by PBE, but is in good accordance with the recent experimental results. Furthermore, the calculated vibron frequencies of these insulating phases agree qualitatively with the experimental infrared (IR) data, revealing that $LiH_9$ could compose the major bulk of the synthesized polyhydrides, with very tiny amount of metastable $LiH_6$ or $LiH_8$.

Our calculation completely revised the phase stability of lithium polyhydrides, and brought the state-of-the-art *ab initio* predictions for the electronic properties and vibron frequencies of lithium polyhydrides to a high level with a much better agreement with the experiment than the previous theoretical assessment. It is worthwhile to note that the hydrogen content in $LiH_9$ reaches a significant 57 wt %, about one order higher than the industrial use requirements. Knuo's experiment hinted that this polyhydride could be synthesized at a much low pressure and high temperature[23], indicative of the attractive prospect of lithium polyhydrides in the application of hydrogen storage.

**Supporting Information Available:**

Phase transitions in solid $H_2$; formation enthalpies ($\Delta H$) of solid $LiH_n$ ($n$ = 2-10, 13) with zero-point energies (ZPE) at 150 GPa; convex hull of $Li_nH$ ($n$ =3-9) at 90 GPa and $NaH_x$ ($x$ =3, 6-12) at 50, 100, and 300 GPa; total and projected density of states (DOS), and phonon density of states (PHDOS) of $LiH_n$ ($n$ = 1, 2-10); shortest bond length of $LiH_n$ ($n$ = 2, 6-10) and $NaH_x$ ($x$ = 6-12); structure details and electron localization functions of Li-H system.

**Acknowledgments**

This work is supported by the National Natural Science Foundation of China under Grant Nos. 11672274, 11274281, and 11174214, the CAEP Research Project under Grant No 2015B0101005, the NSAF under Grant No. U1430117, the Science Challenge project (Grant No. JCKY 2016212A501), and the Fund of National Key Laboratory of Shock Wave and Detonation Physics of China under Grant No. 6142A03010101.

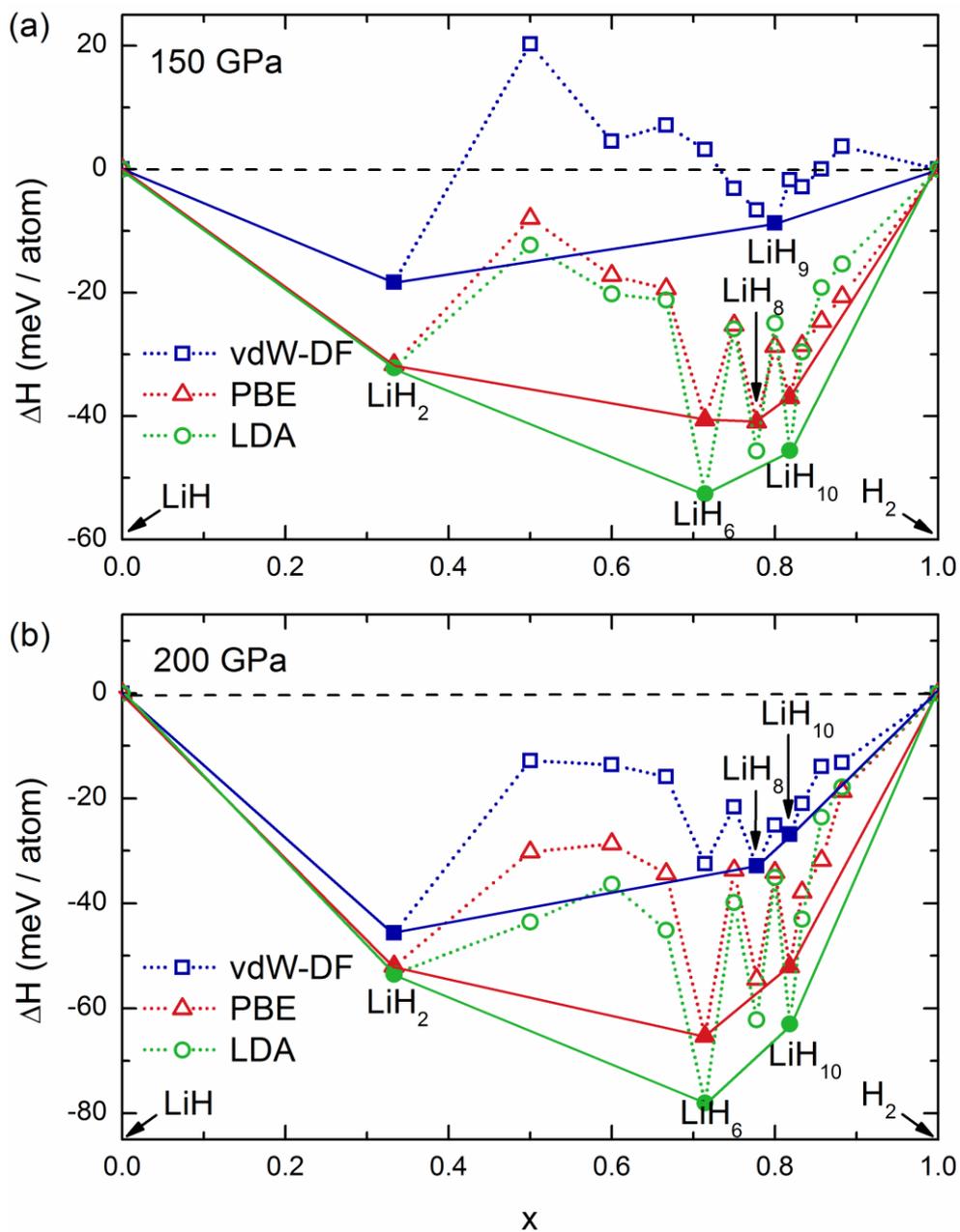

**Figure 1.** Formation enthalpies (ΔH) of solid LiH$_n$ ($n$ = 2-11, 13, 16) with respect to decomposition into LiH and H$_2$ using the vdW-DF, PBE, and LDA functionals at (a) 150 and (b) 200 GPa, respectively. The abscissa $x$ is the fraction of H$_2$ in the structures, and the filled symbols located on the convex hull (the solid lines) represent the stable species against any type of decomposition. The B1 phase of LiH[65], and $P6_3/m$ and $C2/c$ structures of solid H$_2$[66] are used as the reference states for calculating ΔH. The vdW-DF predicts a sequence of structural transitions in solid H$_2$ at 100-200 GPa (see Figure S1 in the Supporting Information (SI)).

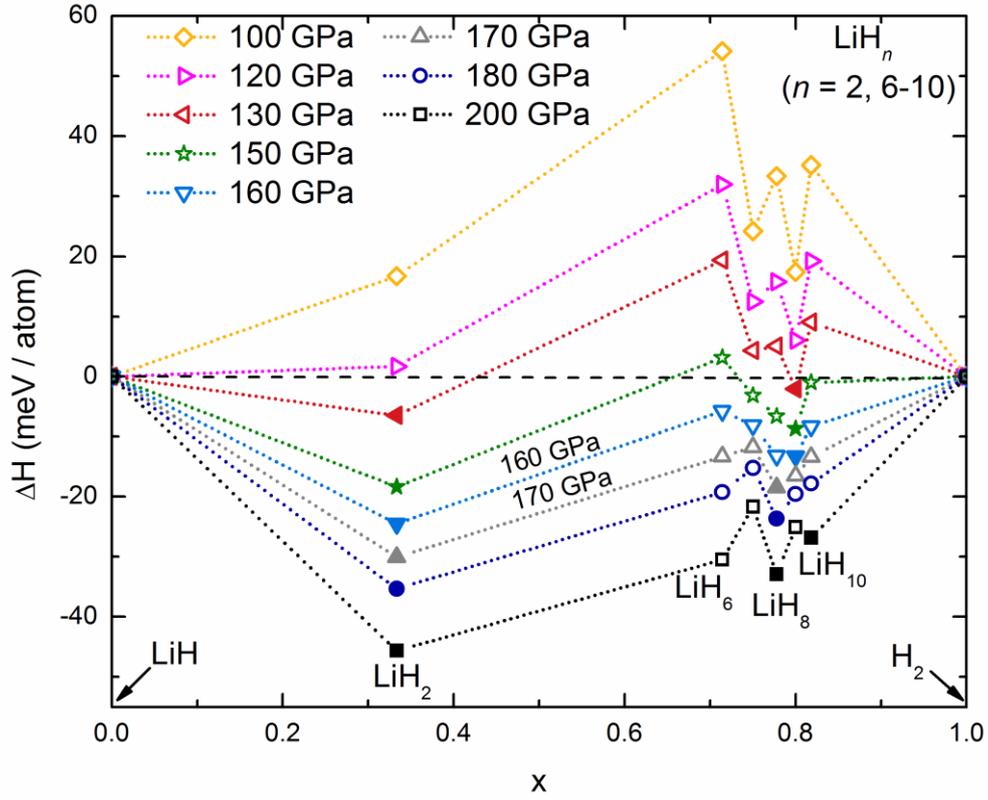

**Figure 2.** Formation enthalpies (ΔH) of solid LiH$_n$ ($n$ = 2, 6-10) with respect to decomposition into LiH and H$_2$ using the vdW-DF functional under various pressures. The abscissa $x$ is the fraction of H$_2$ in the structures, and the filled symbols represent the stable ground-state compounds (on the convex hull) under the corresponding pressure. It can be seen that LiH$_2$ becomes a stable ground-state when above ~120 GPa, LiH$_9$ is stable between 130-170 GPa, and LiH$_8$ and LiH$_{10}$ become stable starting from 180 to 200 GPa, respectively.

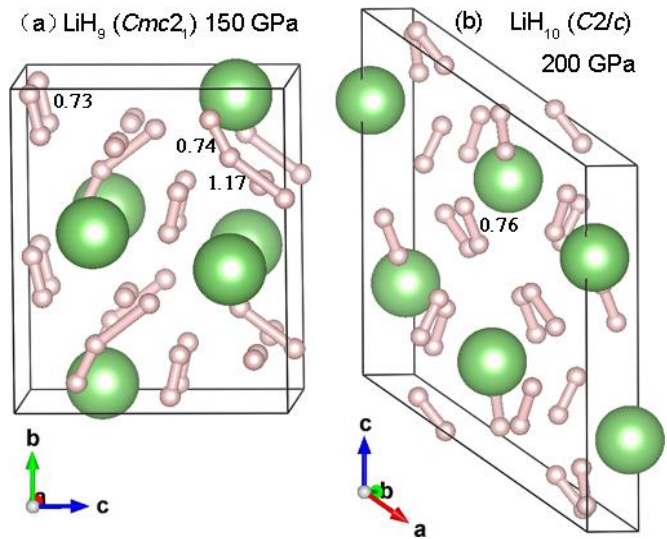

**Figure 3.** Thermodynamic stable structures of (a) LiH$_9$ at 150 GPa and (b) LiH$_{10}$ at 200 GPa, and their bond lengths of H$_2$ units and asymmetric H$_3^-$ clusters.

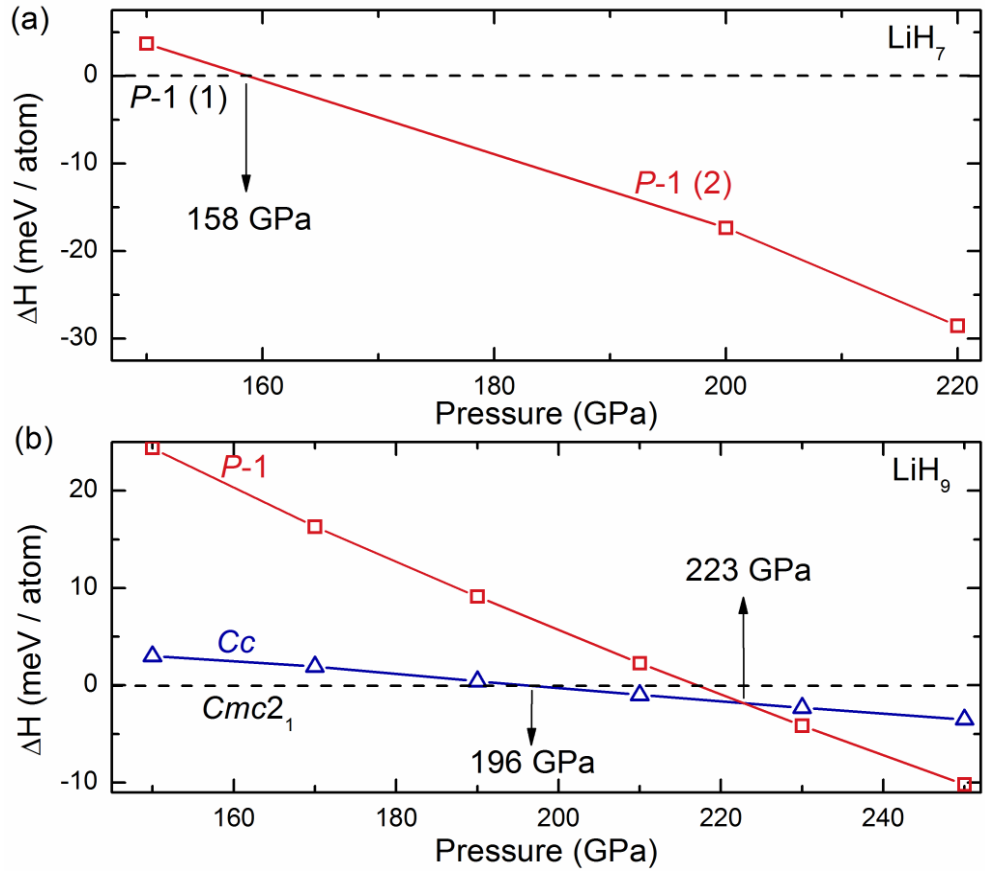

**Figure 4.** Phase transitions of (a) LiH$_7$ and (b) LiH$_9$ predicted by the vdW-DF functional, respectively: Two different structures with the same space group symmetry (*P*-1) appearing in LiH$_7$, and a transition sequence of *Cmc2$_1$*→*Cc*→*P*-1 occurring in LiH$_9$.

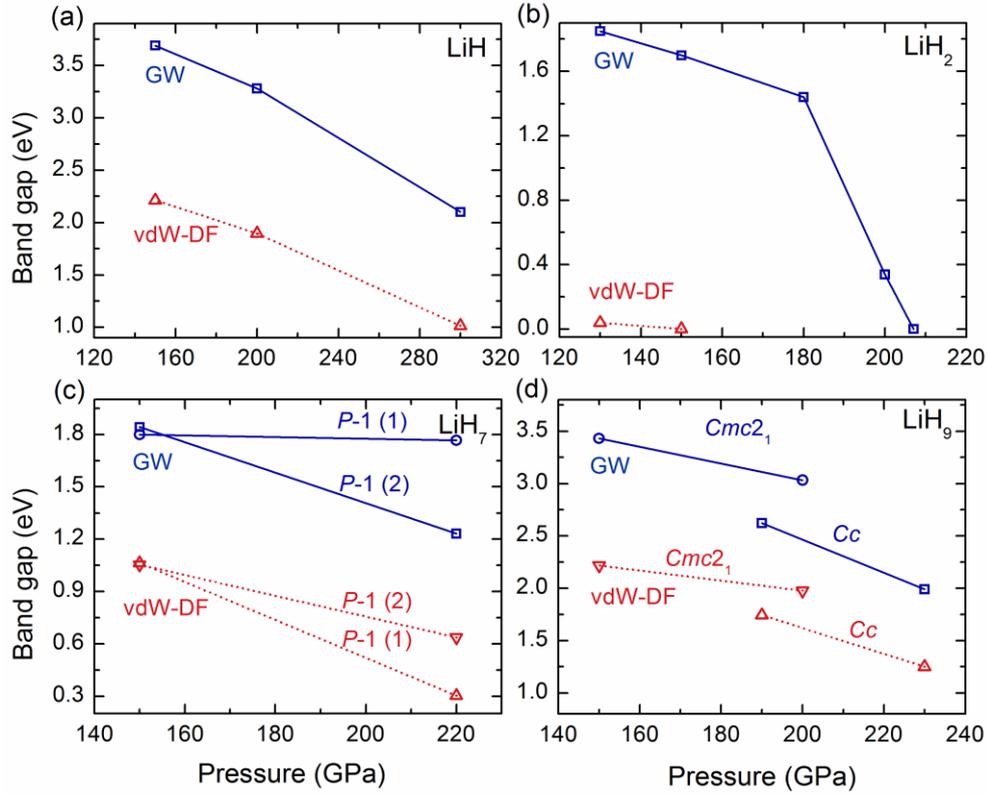

**Figure 5.** Band gaps of (a) LiH (*Fm-3m*), (b) LiH$_2$ (*P4/mbm*), (c) LiH$_7$ (*P*-1(1) and *P*-1(2)), and (d) LiH$_9$ (*Cmc*2$_1$ and *Cc*) calculated with GW method (solid lines) and vdW-DF functional (dotted lines), respectively.

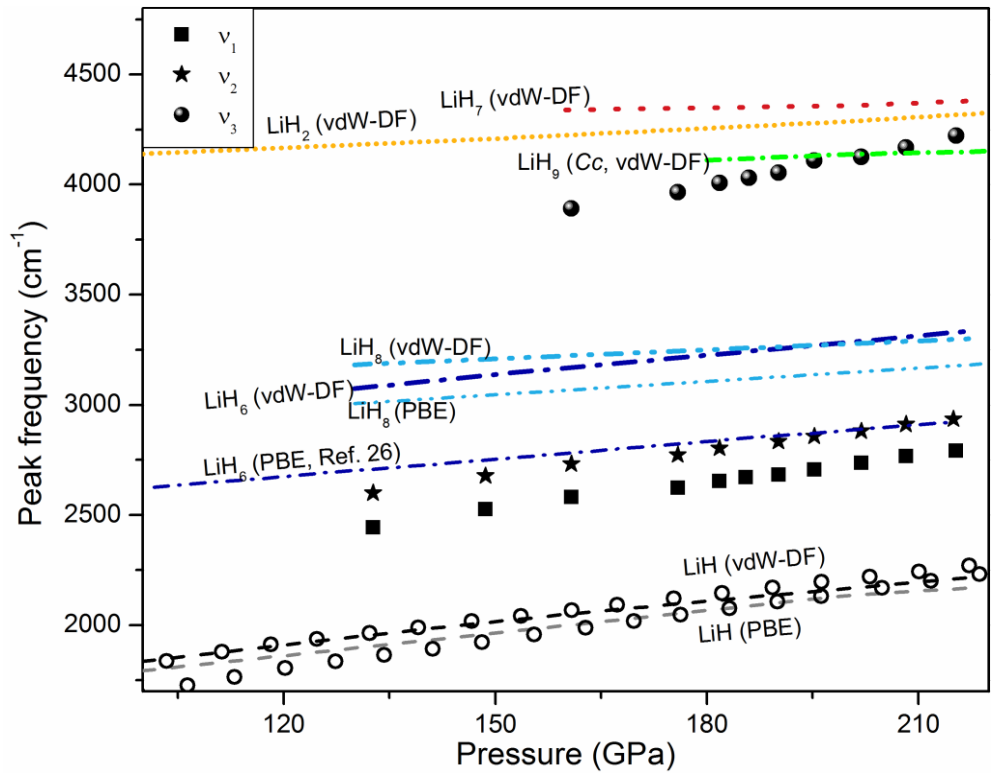

**Figure 6.** Selected vibron frequencies of insulating LiH (*Fm*-3*m*), LiH$_2$ (*P*4/*mbm*), LiH$_7$ (*P*-1(2)), and LiH$_9$ (*Cc*), as well as the metallic LiH$_6$ (*R*-3*m*) and LiH$_8$ (*I*422) calculated with the vdW-DF and PBE functional, respectively. The experimental vibron frequencies ν$_1$, ν$_2$, and ν$_3$ are from Ref. 25. The longitudinal and transverse optical (LO-TO) modes at the *X* point of solid LiH as reported in Ref. 62 are also displayed (open circles).

# SUPPORTING INFORMATION

# Prediction of Stable Ground-State Lithium Polyhydrides under High Pressures

Yangmei Chen,[1,2] Hua Y. Geng,[1,*] Xiaozhen Yan,[1,2,3] Yi Sun,[1] Qiang Wu,[1,*] and Xiangrong Chen[2,*]

[1]*National Key Laboratory of Shock Wave and Detonation Physics, Institute of Fluid Physics, CAEP; P.O. Box 919-102, Mianyang, Sichuan, People's Republic of China, 621900*

[2]*Institute of Atomic and Molecular Physics, College of Physical Science and Technology, Sichuan University, Chengdu, Sichuan, People's Republic of China, 610065*

[3]*School of Science, Jiangxi University of Science and Technology, Ganzhou, Jiangxi, People's Republic of China, 341000*

*Correspondence should be addressed to Hua Y. Geng, Qiang Wu, and Xiangrong Chen (s102genghy@caep.cn, wuqianglsd@163.com, xrchen@scu.edu.cn)

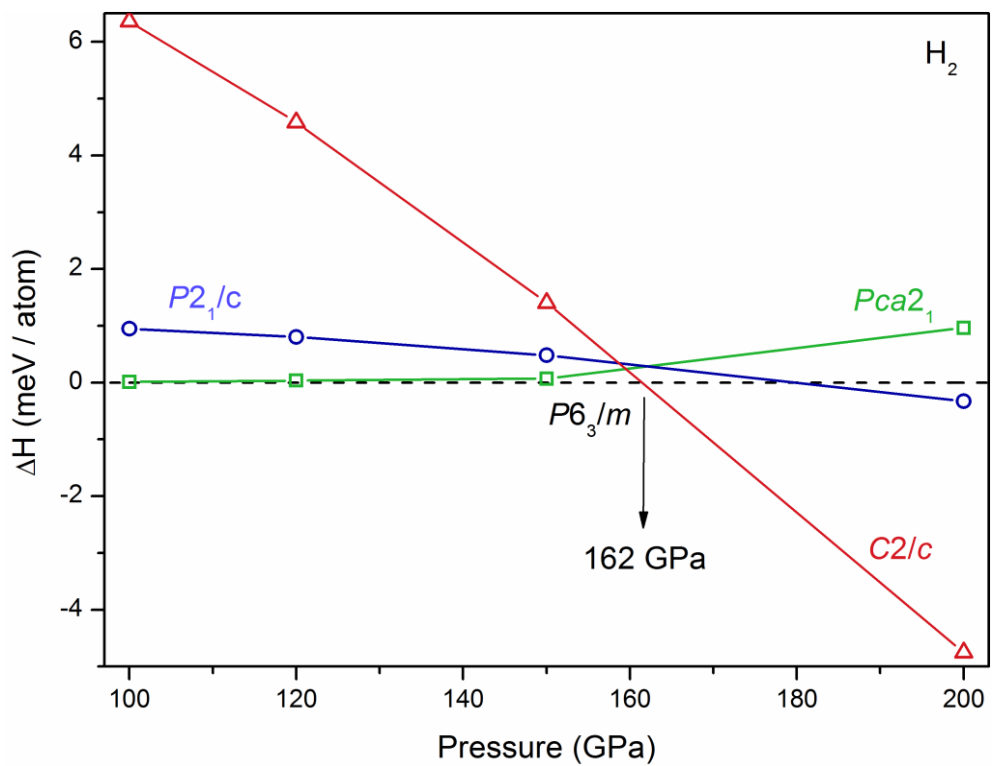

**Figure S1.** Phase transitions in solid $H_2$ calculated with the vdW-DF functional at 100-200 GPa (seeing ref. 26 for beyond 200 GPa). The transition pressure ($P6_3/m \rightarrow C2/c$) is at around 162 GPa.

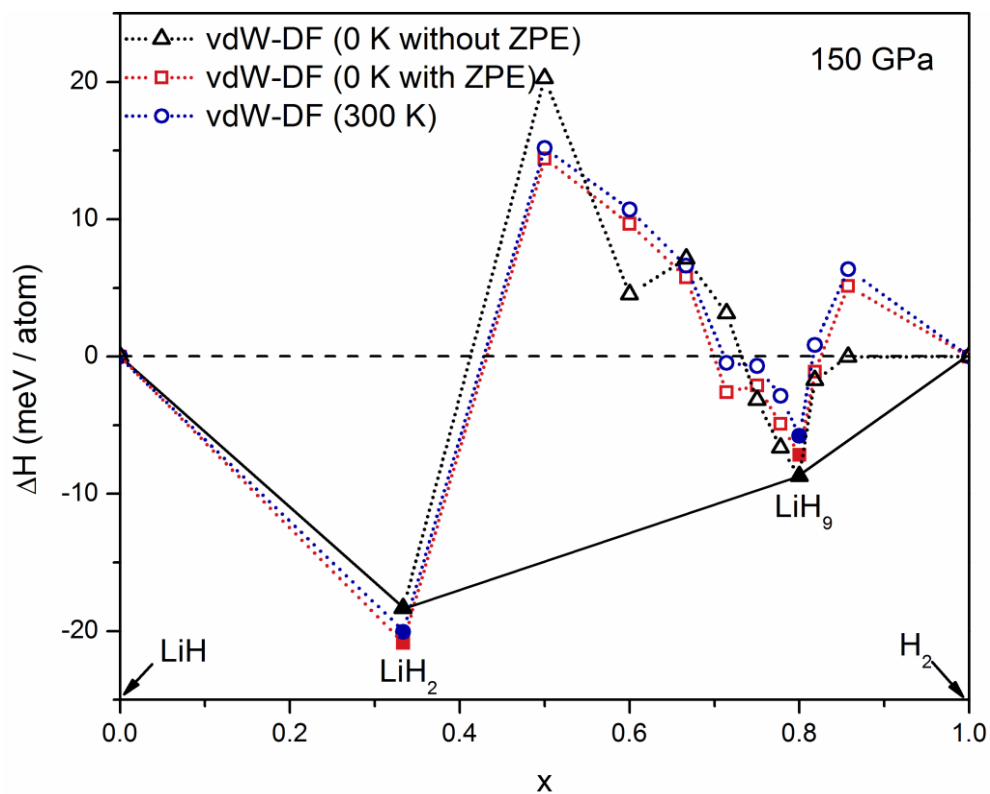

**Figure S2.** Formation enthalpies (ΔH) of solid $LiH_n$ ($n$ = 2-10, 13) with respect to decomposition

into LiH and H$_2$ calculated using the vdW-DF functional at 150 GPa (with and without zero-point energies (ZPE)), and the Gibbs free energy of formation of LiH$_n$ at 300 K, respectively. The abscissa *x* is the fraction of H$_2$ in the structures, and the filled symbols represent the stable ground-state compounds under the corresponding pressure.

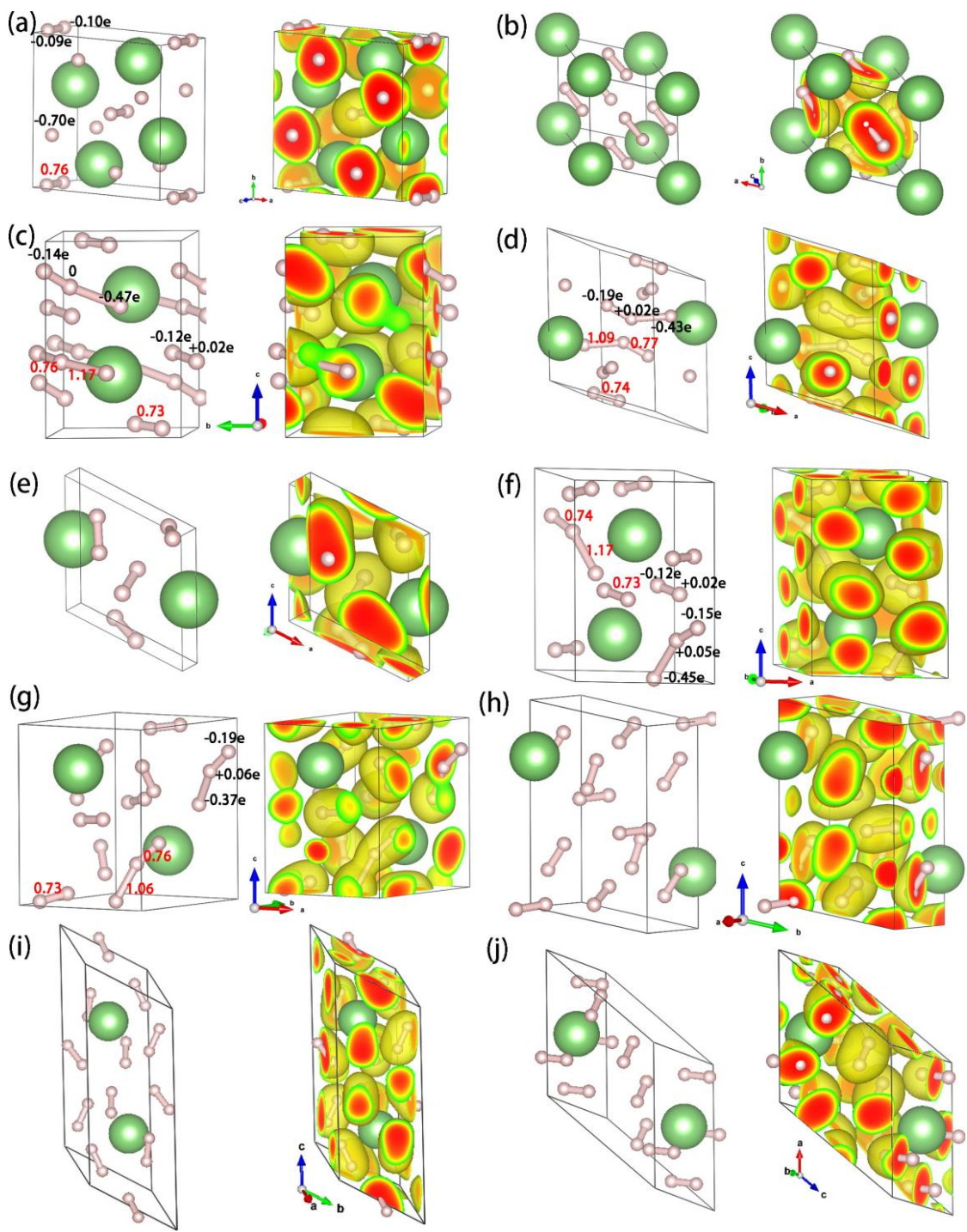

**Figure S3.** Structures and electron localization functions (isosurface = 0.5) of (a) LiH$_2$ (*P*4/*mbm*,

150 GPa), (b) LiH$_6$ (R-3m, 150 GPa), (c) LiH$_7$ (P-1(1), 150 GPa), (d) LiH$_7$ (P-1(2), 200 GPa), (e) LiH$_8$ (I422, 150 GPa) (f) LiH$_9$ (Cmc2$_1$, 150 GPa), (g) LiH$_9$ (Cc, 190 GPa), (h) LiH$_9$ (P-1, 210 GPa), (i) LiH$_{10}$ (C2/c, 200 GPa) and (j) LiH$_{16}$ (I4$_2$m, 150 GPa), respectively.

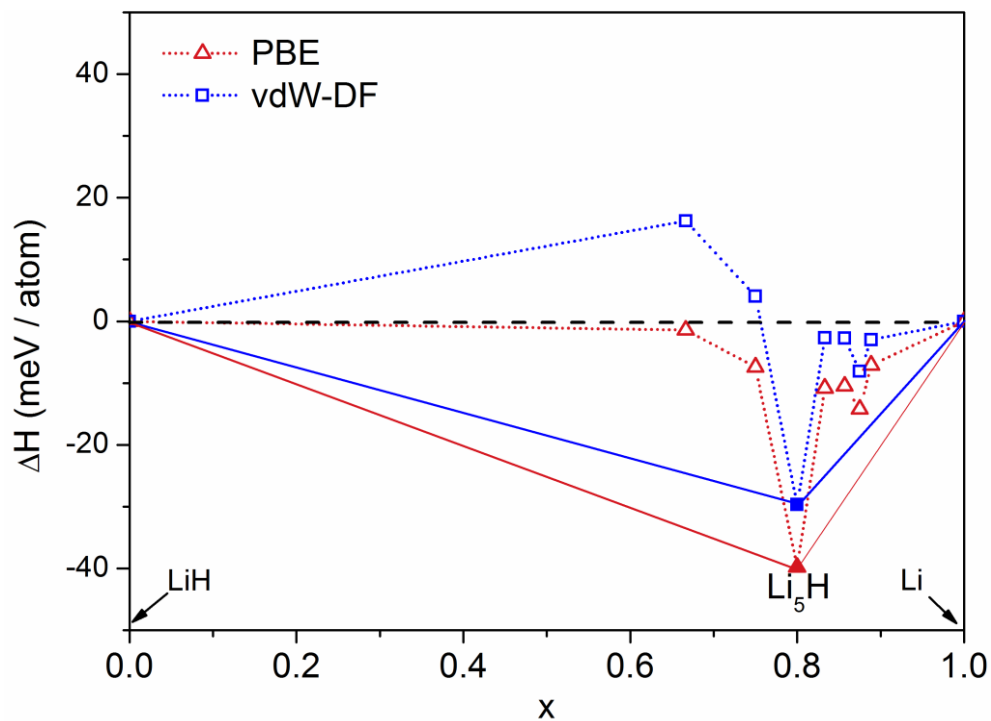

**Figure S4.** Formation enthalpies (ΔH) of Li$_n$H (n =3-9) with respect to the decomposition into Li and LiH using the vdW-DF and PBE density functionals at 90 GPa for the structures reported by Hooper *et al*.[52] The abscissa *x* is the fraction of LiH in the structures, and the filled symbols located on the convex hull (solid lines) represent stable species against any type of decomposition. It is evident the vdW-DF reduces the formation enthalpy of Li$_5$H, but doesn't change the relative phase stability.

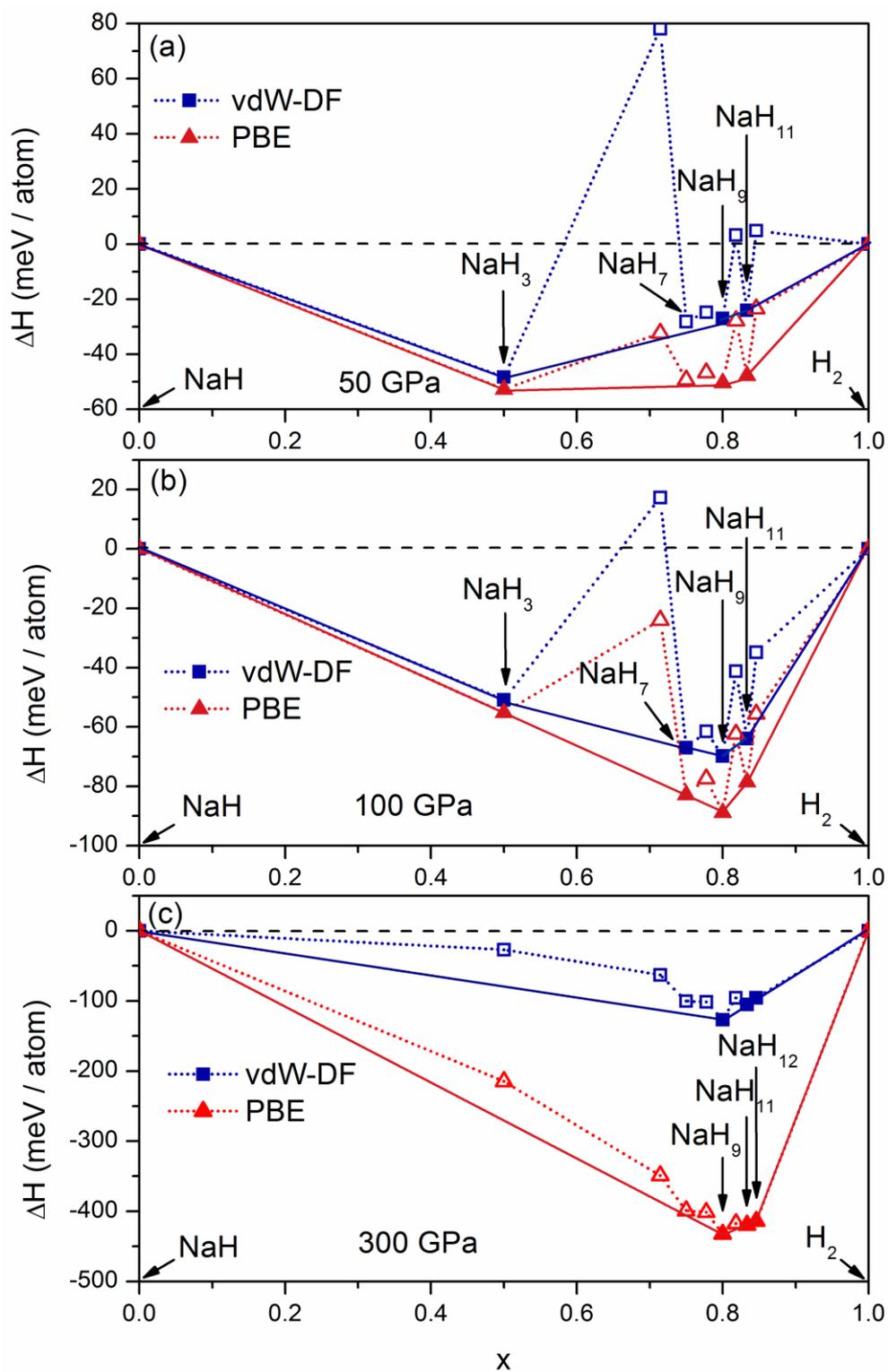

**Figure S5.** Formation enthalpies (ΔH) of NaH$_x$ ($x$ =3, 6-12) with respect to the decomposition into NaH and H$_2$ using the vdW-DF and PBE density functionals for the structures reported in Ref. 24 and 53 at (a) 50 GPa, (b) 100 GPa, and (c) 300 GPa, respectively. The abscissa $x$ is the fraction of H$_2$ in the structures, and the filled symbols located on the convex hull (solid lines) represent stable species against any type of decomposition. It can be seen that van der Waals interaction becomes

important at high pressures, though it still keeps the relative phase stability of PBE.

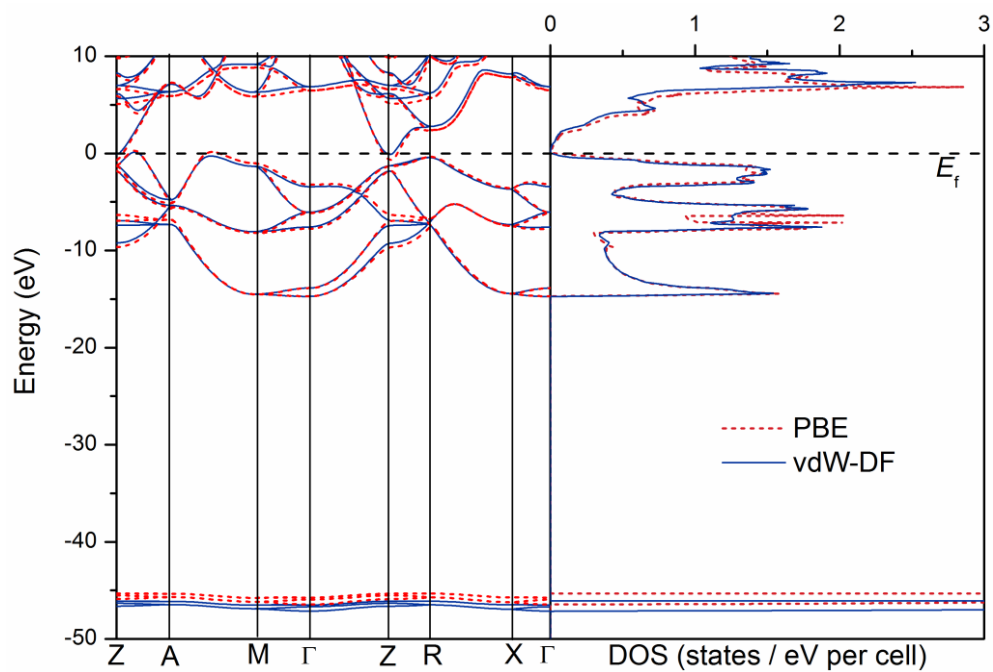

**Figure S6.** Energy band structures and density of states calculated by the vdW-DF (blue solid lines) and PBE (dashed lines in red) functionals for LiH$_2$ at 150 GPa for their respectively optimized structure. The horizontal dashed line indicates the Fermi level.

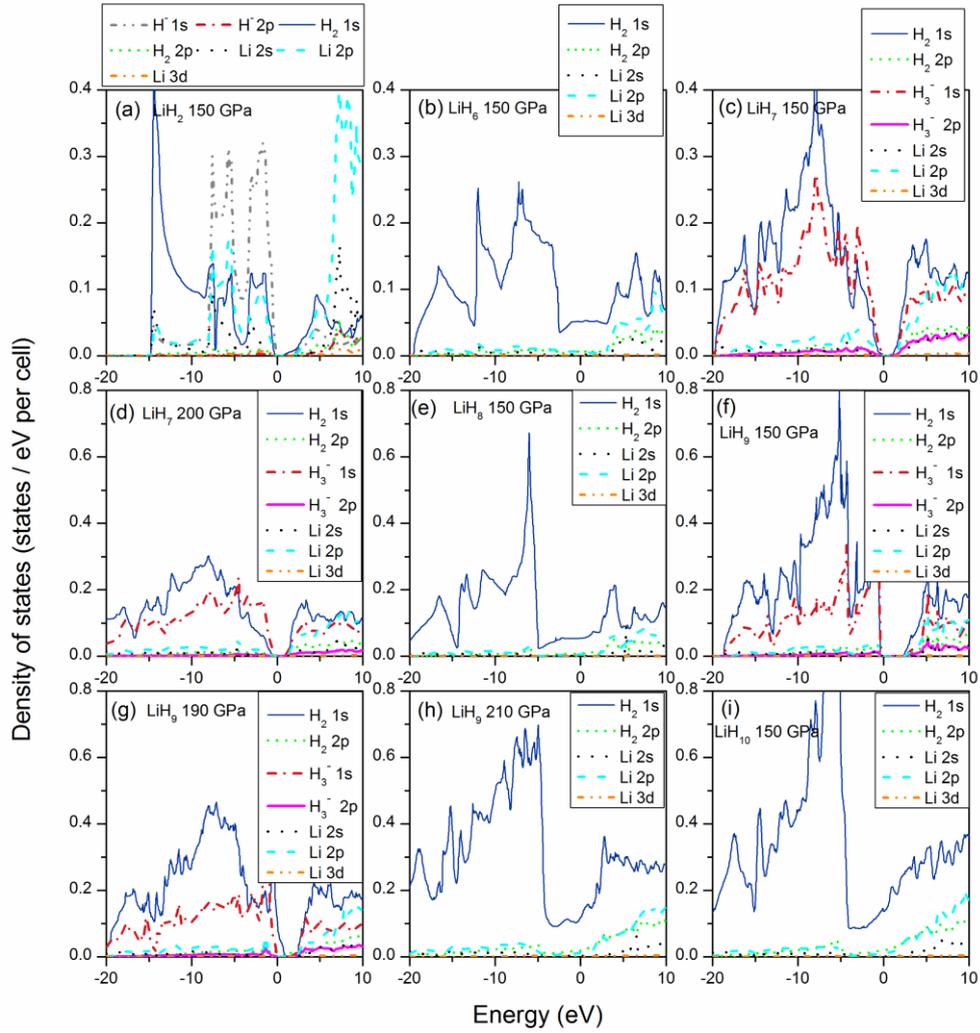

**Figure S7.** Total and projected density of states (DOS) of (a) LiH$_2$ (*P4/mbm*, 150 GPa), (b) LiH$_6$ (*R-3m*, 150 GPa), (c) LiH$_7$ (*P-1*(1), 150 GPa), (d) LiH$_7$ (*P-1*(2), 200 GPa) (e) LiH$_8$ (*I422*, 150 GPa), (f) LiH$_9$ (*Cmc*2$_1$, 150 GPa), (g) LiH$_9$ (*Cc*, 190 GPa), (h) LiH$_9$ (*P-1*, 210 GPa), and (i) LiH$_{10}$ (*C2/c*, 150 GPa) calculated using the vdW-DF functional. The Fermi-level is at zero.

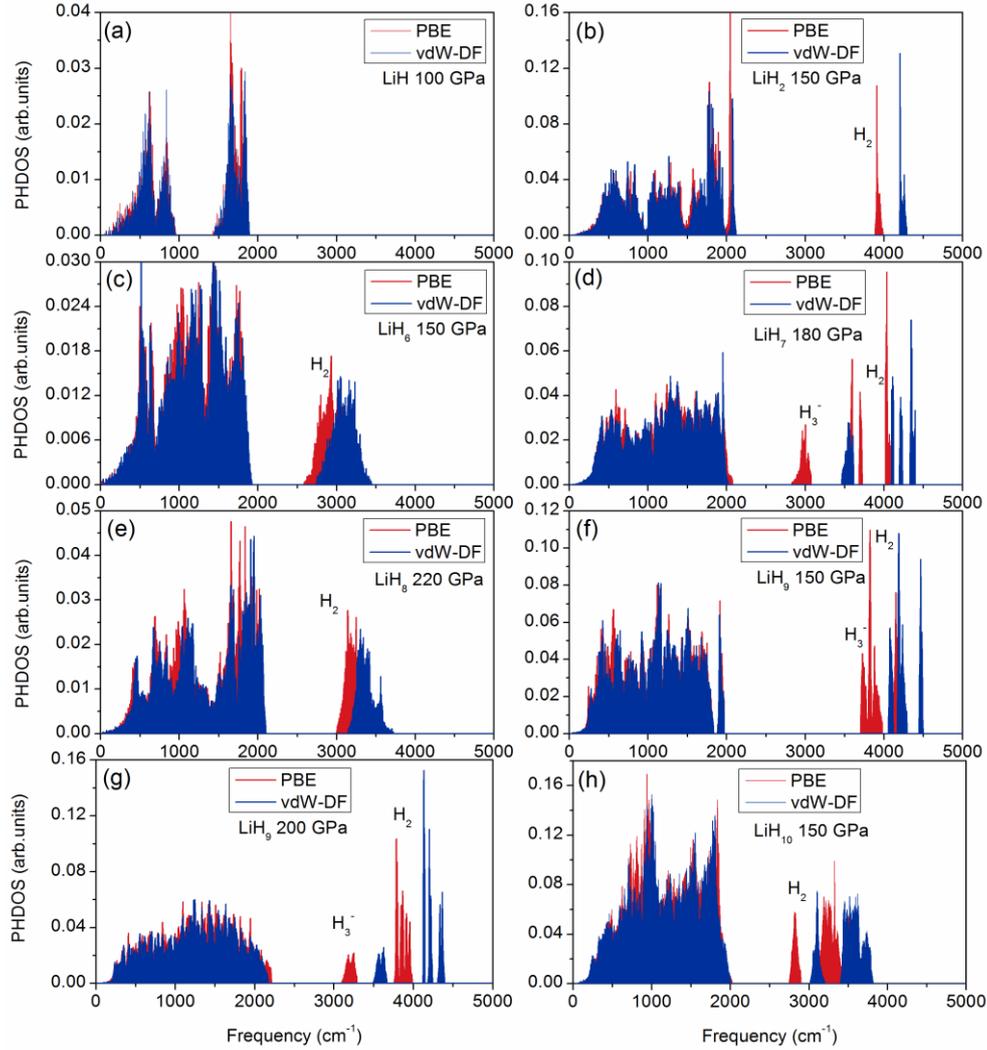

**Figure S8.** The phonon density of states (PHDOS) of (a) LiH (*Fm*-3*m*, 100 GPa), (b) LiH$_2$ (*P4/mbm*, 150 GPa), (c) LiH$_6$ (*R*-3*m*, 150 GPa), (d) LiH$_7$ (*P*-1(2), 180 GPa), (e) LiH$_8$ (*I*422, 220 GPa), (f) LiH$_9$ (*Cmc*2$_1$, 150 GPa), (g) LiH$_9$ (*Cc*, 200 GPa), and (h) LiH$_{10}$ (*C*2/*c*, 150 GPa) calculated using the PBE (red) and vdW-DF (blue) functionals, respectively. Both functionals yield almost the identical low frequency (<2500 cm$^{-1}$) phonon spectra. The frequencies of the vibron modes at the high-frequency regime (>2500 cm$^{-1}$) depend strongly on the choice of the exchange-correlation functional.

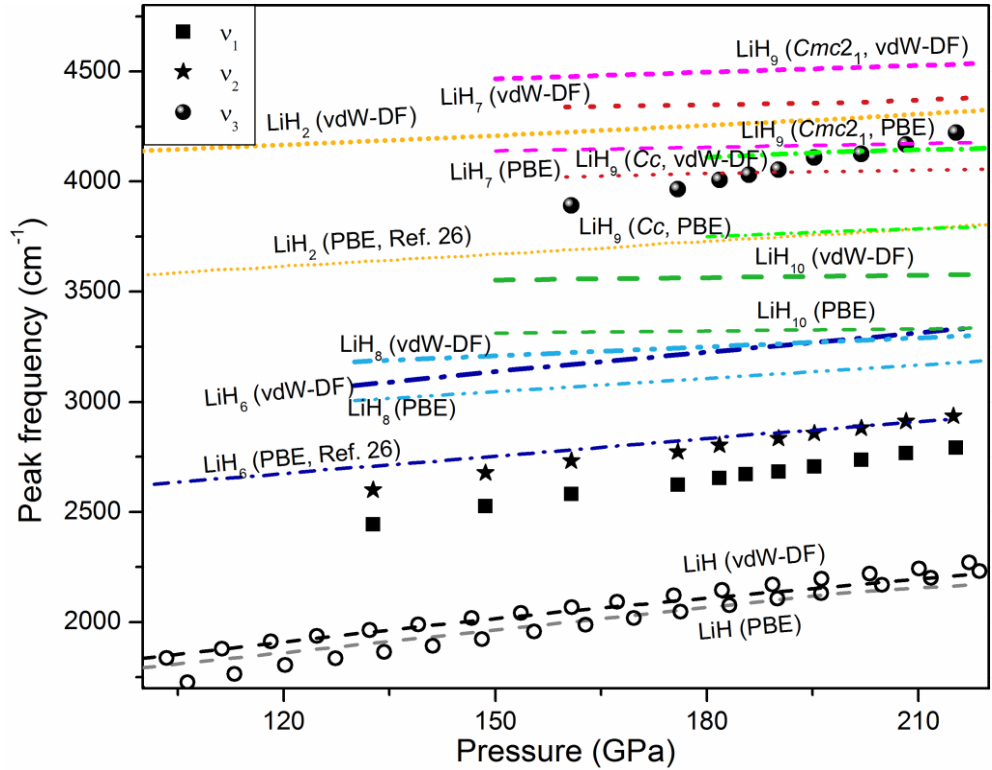

**Figure S9**. Selected vibron frequencies of insulating LiH (*Fm*-3*m*), LiH$_2$ (*P4/mbm*), LiH$_7$ (*P*-1(2)), and LiH$_9$ (*Cmc*2$_1$ and *Cc*), as well as the metallic LiH$_6$ (*R*-3*m*), LiH$_8$ (*I*422), and LiH$_{10}$ (*C*2/*c*) calculated with the vdW-DF and PBE functional, respectively. The experimental vibron frequencies ν$_1$, ν$_2$, and ν$_3$ are from Ref. 24. The longitudinal and transverse optical (LO-TO) modes at the *X* point of solid LiH as reported in Ref. 62 are also displayed (open circles).

**Auxiliary Table S1**. The shortest bond length in LiH$_n$ (*n* = 2, 6-10) at 150 GPa and their relative errors in PBE results with respect to those of the vdW-DF functional.

| LiH$_n$ | H-H (Å) | | | Li-H (Å) | | | Li-Li (Å) | | |
|---|---|---|---|---|---|---|---|---|---|
| 150 GPa | PBE | vdW-DF | Error (%) | PBE | vdW-DF | Error (%) | PBE | vdW-DF | Error (%) |
| LiH$_2$ | 0.762 | 0.734 | 3.8 | 1.500 | 1.508 | -0.5 | 1.797 | 1.801 | -0.2 |
| LiH$_6$ | 0.822 | 0.795 | 3.4 | 1.617 | 1.627 | -0.6 | 2.569 | 2.589 | -0.8 |
| LiH$_7$ | 0.764 | 0.734 | 4.1 | 1.513 | 1.520 | -0.5 | 2.240 | 2.258 | -0.8 |
| LiH$_8$ | 0.809 | 0.784 | 3.2 | 1.517 | 1.527 | -0.7 | 2.915 | 2.967 | -1.8 |
| LiH$_9$ | 0.747 | 0.720 | 3.7 | 1.503 | 1.513 | -0.7 | 2.407 | 2.433 | -1.1 |

| | | | | | | | | |
|---|---|---|---|---|---|---|---|---|
| LiH$_{10}$ | 0.790 | 0.763 | 3.5 | 1.516 | 1.521 | -0.3 | 2.960 | 3.034 | -2.4 |

**Auxiliary Table S2**. The shortest bond length in NaH$_x$ ($x$ = 6-12) at 50 GPa and their relative errors in PBE results with respect to those of the vdW-DF functional.

| NaH$_x$ | H-H (Å) | | | Na-H (Å) | | | Na-Na (Å) | | |
|---|---|---|---|---|---|---|---|---|---|
| 50 GPa | PBE | vdW-DF | Error (%) | PBE | vdW-DF | Error (%) | PBE | vdW-DF | Error (%) |
| NaH$_6$ | 0.754 | 0.751 | 0.4 | 1.914 | 1.959 | -2.3 | 2.817 | 2.971 | -5.2 |
| NaH$_7$ | 0.757 | 0.734 | 3.1 | 1.900 | 1.912 | -0.6 | 2.951 | 2.998 | -1.6 |
| NaH$_8$ | 0.756 | 0.734 | 3.0 | 1.904 | 1.909 | -0.3 | 3.072 | 3.156 | -2.7 |
| NaH$_9$ | 0.760 | 0.734 | 3.5 | 1.954 | 1.964 | -0.5 | 3.068 | 3.145 | -2.4 |
| NaH$_{10}$ | 0.764 | 0.734 | 4.1 | 2.045 | 2.063 | -0.9 | 3.294 | 3.333 | -1.2 |
| NaH$_{11}$ | 0.753 | 0.731 | 3.0 | 1.956 | 1.975 | -1.0 | 3.526 | 3.569 | -1.2 |
| NaH$_{12}$ | 0.757 | 0.740 | 2.3 | 1.954 | 1.976 | -1.1 | 3.380 | 3.399 | -0.6 |

**Auxiliary Table S3**. Lattice parameters of the primitive cell of LiH$_n$ ($n$ = 2, 6-10) at 150 GPa optimized by using the PBE and vdW-DF functional, respectively.

| Phase | Space group | Lattice parameters PBE | Lattice parameters vdW-DF |
|---|---|---|---|
| LiH$_2$ | *P4/mbm* | a = b = 4.087 Å | a = b = 4.115 Å |
| | | c = 1.961 Å | c = 1.967 Å |
| | | α = β = γ = 90.000º | α = β = γ = 90.000º |
| LiH$_6$ | *R-3m* | a = b = c = 2.569 Å | a = b = c = 2.589 Å |
| | | α = β = γ = 74.987º | α = β = γ = 75.084º |
| LiH$_7$ | *P-1* | a = 3.111 Å | a = 3.124 Å |
| | | b = 3.127 Å | b = 3.158 Å |
| | | c = 4.352 Å | c = 4.418 Å |
| | | α = 94.804º | α = 94.981º |
| | | β = 73.305º | β = 72.895º |

| | | | γ = 118.896º | γ = 118.583º |
|---|---|---|---|---|
| LiH$_8$ | I422 | | a = b = c = 2.959 Å | a = b = c = 2.967 Å |
| | | | α = 103.758º | α = 103.669º |
| | | | β = γ = 112.729º | β = γ = 112.448º |
| LiH$_9$ | Cmc2$_1$ | | a = b = 3.256 Å | a = b = 3.291 Å |
| | | | c = 4.649 Å | c = 4.694 Å |
| | | | α = β = 90.000º | α = β = 90.000º |
| | | | γ = 118.887º | γ = 118.858º |
| LiH$_{10}$ | C2/c | | a = b = 3.065 Å | a = b = 3.071 Å |
| | | | c = 6.830 Å | c = 6.985 Å |
| | | | α = β = 119.029º | α = β = 118.986º |
| | | | γ = 60.443º | γ = 61.347º |

**Auxiliary Table S4.** Structure details of the conventional unit cell of LiH$_n$ ($n$ = 2, 6-10) obtained using the vdW-DF functional.

| Phase | Pressure (GPa) | Space group | Lattice parameters | Atomic coordinates |
|---|---|---|---|---|
| LiH$_2$ | 150 | P4/mbm | a = b = 4.115 Å | H1(4e) 0.0000 0.0000 0.3135 |
| | | | c = 1.967 Å | H2(4g) 0.8509 0.6490 0.0000 |
| | | | α = β = γ = 90.000º | Li(4h) 0.8453 0.3453 0.5000 |
| LiH$_6$ | 150 | R-3m | a = b = c = 2.589 Å | H(6h) -0.9334 -0.4253 -0.4253 |
| | | | α = β = γ = 75.084º | Li(1a) 0.0000 0.0000 0.0000 |
| LiH$_7$ | 150 | P-1 | a = 3.124 Å | H1(2i) 0.9235 0.9800 0.1552 |
| | | | b = 3.158 Å | H2(2i) 0.6086 0.8858 0.5884 |
| | | | c = 4.418 Å | H3(2i) 0.5268 0.2110 0.0204 |
| | | | α = 94.981 º | H4(2i) 0.1560 0.9059 0.3791 |
| | | | β = 72.895º | H5(2i) 0.1422 0.8383 0.7611 |
| | | | γ = 118.583º | H6(2i) 0.2505 0.5634 0.9683 |
| | | | | H7(2i) 0.8387 0.5295 0.3380 |
| | | | | Li(2i) 0.3632 0.5472 0.3041 |

| | | | | | |
|---|---|---|---|---|---|
| LiH$_7$ | 200 | P-1 | a = 3.039 Å | H1(2i) 0.9841 0.7061 0.3294 | |
| | | | b = 3.070 Å | H2(2i) 0.5776 0.0313 0.2073 | |
| | | | c = 4.285 Å | H3(2i) 0.9281 0.3057 0.9681 | |
| | | | α = 108.962º | H4(2i) 0.3044 0.7103 0.0293 | |
| | | | β = 107.153º | H5(2i) 0.5817 0.7192 0.3818 | |
| | | | γ = 60.606º | H6(2i) 0.6419 0.2946 0.5770 | |
| | | | | H7(2i) 0.3757 0.2700 0.1670 | |
| | | | | Li(2i) 0.9822 0.8142 0.6934 | |
| LiH$_8$ | 150 | I422 | a = b = 3.299 Å | H(16k) 0.3739 -0.2095 0.3428 | |
| | | | c = 3.666º | Li(2b) 0.0000 0.0000 0.5000 | |
| | | | α = β = γ = 90.000º | | |
| LiH$_9$ | 150 | Cmc2$_1$ | a = 3.347 Å | H1(8b) 0.7598 -0.1614 -0.9082 | |
| | | | b = 5.667 Å | H2(8b) 0.7672 0.0354 -0.4436 | |
| | | | c = 4.694 Å | H3(8b) 0.6074 -0.3272 -0.1152 | |
| | | | α = β = γ = 90.000º | H4(4a) 0.5000 0.4048 -0.2709 | |
| | | | | H5(4a) 0.5000 0.1616 -0.0025 | |
| | | | | H6(4a) 0.5000 0.2884 -0.2001 | |
| | | | | Li(4a) 0.5000 0.0567 -0.7061 | |
| LiH$_9$ | 190 | Cc | a = 3.184 Å | H1(4a) -0.9718 -0.1325 -0.7764 | |
| | | | b = 5.572 Å | H2(4a) -0.6324 0.0649 -0.8454 | |
| | | | c = 5.575 Å | H3(4a) -0.8940 -0.0832 -0.6365 | |
| | | | α = γ = 90.000º | H4(4a) -0.1027 0.0668 -0.9548 | |
| | | | β = 125.202º | H5(4a) -0.3311 -0.1160 -0.4712 | |
| | | | | H6(4a) -0.6188 0.1176 -0.5526 | |
| | | | | H7(4a) -0.3479 -0.0920 -0.2285 | |
| | | | | H8(4a) -0.0176 0.1592 -0.5298 | |
| | | | | H9(4a) -0.4028 -0.1954 -0.0159 | |
| | | | | Li(4a) -0.4686 -0.1989 -0.7592 | |
| LiH$_9$ | 210 | P-1 | a = 2.909 Å | H1(2i) 0.3962 0.8838 0.6905 | |

| | | | | |
|---|---|---|---|---|
| | | | b = 2.989 Å | H2(2i) 0.5314 0.9452 0.1630 |
| | | | c = 5.206 Å | H3(2i) 0.4394 0.5192 0.8493 |
| | | | α = 102.937º | H4(2i) 0.8424 0.5005 0.7664 |
| | | | β = 92.369º | H5(2i) 0.0974 0.5533 0.3769 |
| | | | γ = 117.981º | H6(2i) 0.2801 0.3564 0.5682 |
| | | | | H7(2i) 0.7381 0.4298 0.0595 |
| | | | | H8(2i) 0.3677 0.1471 0.5268 |
| | | | | H9(2i) 0.9470 0.1022 0.0237 |
| | | | | Li(2i) 0.9476 0.0158 0.7281 |
| LiH$_{10}$ | 150 | $C2/c$ | a = 5.283 Å | H1(8f) -0.9514 -0.7058 -1.5528 |
| | | | b = 3.134 Å | H2(8f) -0.9087 -0.2054 -1.1327 |
| | | | c = 6.985 Å | H3(8f) -0.8215 -0.8986 -1.3448 |
| | | | α = γ = 90.000º | H4(8f) -0.8032 -0.7294 -0.9774 |
| | | | β = 124.293º | H5(8f) -0.2397 -0.9585 -0.2811 |
| | | | | Li(4e) -0.5000 -0.8537 -0.7500 |

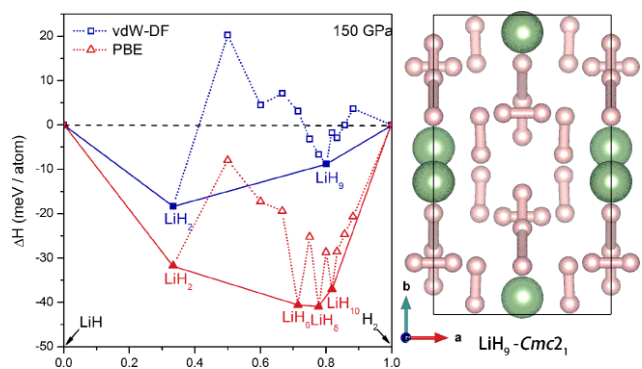

Formation enthalpies (ΔH) of solid LiH$_n$ ($n$ = 2-11, 13, 16) with respect to decomposition into LiH and H$_2$ using the vdW-DF and PBE functionals at 150 GPa, and the schematic depiction of the $Cmc2_1$ structure of LiH$_9$ at 150 GPa.